\documentclass[twocolumn, tighten, times]{aastex631}
\usepackage{graphicx}
\usepackage{epstopdf}
\usepackage{float} 
\usepackage{bm}

\usepackage{color}
\usepackage{soul}

\shorttitle{Classification of COSMOS}
\shortauthors{Song. ET AL}
\graphicspath{{./}{figures/}}

\begin{document}

\title{USmorph: An Updated Framework of Automatic Classification of Galaxy Morphologies and Its Application to Galaxies in the COSMOS Field}

\author[0000-0002-0846-7591]{Jie Song}
\affil{Deep Space Exploration Laboratory / Department of Astronomy, University of Science and Technology of China, Hefei 230026, People’s Republic of China; \url{xkong@ustc.edu.cn}} 
\affil{School of Astronomy and Space Science, University of Science and Technology of China, Hefei 230026, People’s Republic of China}

\author[0000-0001-9694-2171]{GuanWen Fang}
\altaffiliation{GuanWen Fang and Jie Song contributed equally to this work}
\affil{Institute of Astronomy and Astrophysics, Anqing Normal University, Anqing 246133, People's Republic of China; \url{wen@mail.ustc.edu.cn}}

\author{Shuo Ba}
\affil{School of Engineering, Dali University, Dali 671003, People's Republic of China}

\author[0000-0001-8078-3428]{Zesen Lin}
\affil{Department of Physics, The Chinese University of Hong Kong, Shatin, N.T., Hong Kong S.A.R., People’s Republic of China}

\author[0000-0003-3196-7938]{Yizhou Gu}
\affil{School of Physics and Astronomy, Shanghai Jiao Tong University, 800 Dongchuan Road, Minhang, Shanghai 200240, People’s Republic of China} 

\author[0000-0002-5133-2668]{Chichun Zhou}
\affil{School of Engineering, Dali University, Dali 671003, People's Republic of China}

\author[0000-0002-2504-2421]{Tao Wang}
\affil{School of Astronomy and Space Science, Nanjing University, Nanjing, Jiangsu 210093, People's Republic of China}
\affil{Key Laboratory of Modern Astronomy and Astrophysics, Nanjing University, Ministry of Education, Nanjing 210093, People's Republic of China }

\author[0000-0002-0901-9328]{Cai-Na Hao}
\affil{Tianjin Astrophysics Center, Tianjin Normal University, Tianjin 300387, People's Republic of China}

\author[0000-0003-2390-7927]{ Guilin Liu}
\affil{Key Laboratory for Research in Galaxies and Cosmology, Department of Astronomy, University of Science and
Technology of China, Hefei, Anhui 230026, People's Republic of China}
\affil{School of Astronomy and Space Science, University of Science and Technology of China, Hefei 230026, People’s Republic of China}

\author[0000-0003-1632-2541]{Hongxin Zhang}
\affil{Key Laboratory for Research in Galaxies and Cosmology, Department of Astronomy, University of Science and
Technology of China, Hefei, Anhui 230026, People's Republic of China}
\affil{School of Astronomy and Space Science, University of Science and Technology of China, Hefei 230026, People’s Republic of China}

\author[0000-0002-6873-8779]{Yao Yao}
\affil{Deep Space Exploration Laboratory / Department of Astronomy, University of Science and Technology of China, Hefei 230026, People’s Republic of China; \url{xkong@ustc.edu.cn}} 
\affil{School of Astronomy and Space Science, University of Science and Technology of China, Hefei 230026, People’s Republic of China}

\author[0000-0002-7660-2273]{Xu Kong}
\affil{Deep Space Exploration Laboratory / Department of Astronomy, University of Science and Technology of China, Hefei 230026, People’s Republic of China; \url{xkong@ustc.edu.cn}} 
\affil{School of Astronomy and Space Science, University of Science and Technology of China, Hefei 230026, People’s Republic of China}

\begin{abstract}
Morphological classification conveys abundant information on the formation, evolution, and environment of galaxies. In this work, we refine the two-step galaxy morphological classification framework ({\tt\string USmorph}), which employs a combination of unsupervised machine learning (UML) and supervised machine learning (SML) techniques, along with a self-consistent and robust data preprocessing step. The updated method is applied to the galaxies with $I_{\rm mag}<25$ at $0.2<z<1.2$ in the COSMOS field. Based on their HST/ACS I-band images, we classify them
into five distinct morphological types: spherical (SPH, 15,200), early-type disk (ETD, 17,369), late-type disk (LTD, 21,143), irregular disk (IRR, 28,965), and unclassified (UNC, 17,129). In addition, we have conducted both parametric and nonparametric morphological measurements. For galaxies with stellar masses exceeding $10^{9}M_{\sun}$, a gradual increase in effective radius from SPHs to IRRs is observed, accompanied by a decrease in the S\'{e}rsic index. Nonparametric morphologies reveal distinct distributions of galaxies across the $Gini-M_{20}$ and $C-A$ parameter spaces for different categories. Moreover, different categories exhibit significant dissimilarity in their $G_2$ and $\Psi$ distributions. We find morphology to be strongly correlated with redshift and stellar mass. The consistency of these classification results with expected correlations among multiple parameters underscores the validity and reliability of our classification method, rendering it a valuable tool for future studies.
\end{abstract}

\keywords{Galaxy structure (622), Astrostatistics techniques (1886), Astronomy data analysis (1858)}

\section{Introduction} \label{sec:intro}

In the realm of observational cosmology, galaxy morphology stands out as one of the most readily accessible properties, intimately intertwined with numerous other physical attributes (such as color, gas content, star formation rate, stellar mass, and environment) of galaxies (e.g., \citealt{kauffmannEnvironmentalDependenceRelations2004, omandConnectionGalaxyStructure2014, schawinskiGreenValleyRed2014, kawinwanichakijEffectLocalEnvironment2017,guMorphologicalEvolutionAGN2018, lianouDustPropertiesStar2019}). It can offer insights into the evolutionary history and assembly processes of galaxies. For example, as galaxies evolve, their morphological features may transition towards being more bulge-dominated and compact (e.g., \citealt{martigMORPHOLOGICALQUENCHINGSTAR2009, dimauroCoincidenceMorphologyStar2022}). Consequently, accurately estimating galaxy morphology at each epoch within the universe is of fundamental importance for unraveling the intricate tapestry of galaxy evolution.

Several methods are currently available for characterizing galaxy morphology,
with one of the most direct being visual inspection. In a pioneering study, \cite{Hubble_1926} systematically analyzed approximately 400 galaxies and categorized them using what is now known as the ``Hubble tuning fork'' classification. Furthermore, the Galaxy Zoo project engaged the efforts of nearly half a million volunteers, resulting in the morphological classification of over one million galaxies (e.g., \citealt{lintottGalaxyZooData2011, simmonsGalaxyZooQuantitative2017, 2017MNRAS.464.4176W, walmsleyGalaxyZooDECaLS2021}). These citizen science initiatives have significantly expanded our knowledge of galaxy morphologies. However, this method is characterized by low efficiency, high cost, and inherent subjective bias, which may not be suitable for future large-sky surveys.

Besides visual inspection, galaxy morphological classifications can be also obtained with several features from the raw images. This can be achieved through parametric measurement, which involves modeling the galaxy's light distribution with an analytic function (e.g., \citealt{1948AnAp...11..247D, 1963BAAA....6...41S, 1970ApJ...160..811F}), and nonparametric measurement (e.g., \citealt{conseliceAsymmetryGalaxiesPhysical2000, conseliceRelationshipStellarLight2003, lotzNewNonparametricApproach2004, lotzEvolutionGalaxyMergers2008, conseliceEvolutionGalaxyStructure2014a}). Using Principal Component Analysis techniques and based on structural parameters (e.g., C, A, Gini, $M_{\rm 20}$, ellipticity, and S\'{e}rsic index), \cite{2007ApJS..172..406S} classified approximately 56,000 galaxies in the COSMOS (Cosmic Evolution Survey, \citealt{scovilleCosmicEvolutionSurvey2007}) field into early-type, disk, and irregular galaxies. Similarly, according to the distribution of galaxies in the morphological structural parameter space, \cite{2007ApJS..172..270C} and \cite{2009A&A...503..379T} also obtained the morphological classification results in the COSMOS field. This approach represents the characteristics of galaxies using a few parameters, reducing the complexity of describing galaxy morphology, but it also rejects the rich information hidden beneath all the pixels, which may lead to misclassification in some cases.

In recent years, the rapid advancement of computer technology has enabled the application of machine-learning techniques for galaxy morphological classification.
This include feature-based methods (e.g., \citealt{2018MNRAS.474.5232S, 2007A&A...468..937H, 2010MNRAS.406..342B, 2010arXiv1005.0390G}) and image-based methods (e.g., \citealt{2019Ap&SS.364...55Z, 2020ApJ...895..112G, vega-ferreroPushingAutomatedMorphological2021}).
Convolutional Neural Networks (CNNs, \citealt{SCHMIDHUBER201585}), in particular, have demonstrated their ability to replicate human perception and have been successfully used to determine galaxy morphological types in both the local universe (e.g., SDSS - Sloan Digital Sky Survey, \citealt{2000AJ....120.1579Y}, DES - Dark Energy Survey, \citealt{2016MNRAS.460.1270D}) and the high-redshift universe (e.g., CANDELS - Cosmic Assembly Near-infrared Deep Extragalactic Legacy Survey, \citealt{2011ApJS..197...35G}) (e.g., \citealt{dielemanRotationinvariantConvolutionalNeural2015a, 2019Ap&SS.364...55Z, 2020ApJ...895..112G, vega-ferreroPushingAutomatedMorphological2021, 2021MNRAS.506..659C}). However, as an SML method, CNNs require large pre-labeled datasets as training sets, which are typically obtained through visual inspection. As mentioned earlier, this step can be time-consuming and negates some of the advantages of machine learning.

UML offers an alternative approach for galaxy morphological classification, eliminating the need for pre-labeled training sets. This characteristic renders it suitable for morphological analysis in the context of large-scale surveys (e.g., \citealt{2015A&C....12...60S, ralphRadioGalaxyZoo2019}). Typically, UML methods involve two key steps: (1) feature extraction from raw images and (2) clustering galaxies with similar features. Numerous UML methods have already found application in various studies (e.g., \citealt{hockingAutomaticTaxonomyGalaxy2018, 2022icec.confE...1F,2020MNRAS.491.1408M, chengHubbleSequenceExploring2021b}). For instance, using a sample from the SDSS, \cite{chengHubbleSequenceExploring2021b} has demonstrated the effectiveness of this UML approach for galaxy morphological classification, providing a robust scheme.

However, many UML methods only focus on one single clustering algorithm, potentially leading to inconsistent clustering outcomes when different similarity definitions are employed, which may result in misclassification. To obtain morphological classifications with high confidence, \cite{zhouAutomaticMorphologicalClassification2022} introduced a Bagging-based multi-clustering model that incorporates three diverse clustering algorithms. This approach, combined with convolutional autoencoding (CAE; \citealt{masseyPixelbasedCorrectionCharge2010}) for feature extraction from images, yielded reliable classification results at the cost of excluding disputed sources with inconsistent voting. Subsequently, \cite{fangAutomaticClassificationGalaxy2023} used the classification results from \cite{zhouAutomaticMorphologicalClassification2022} as a training set in an SML algorithm to determine the morphological types of previously rejected sources from the same dataset. This framework that combines \textbf{U}ML and \textbf{S}ML methods for \textbf{morph}ological classification (which is named as {\tt\string USmorph}) can help us obtain reliable and complete galaxy morphological type efficiently, which makes it suitable for future large-field sky surveys, such as the ones from CSST (Chinese Space Station Telescope, \citealt{2011SSPMA..41.1441Z, 2018cosp...42E3821Z}), JWST (James Webb Space Telescope, \citealt{2006SSRv..123..485G}), Euclid Space Telescope \citep{2011arXiv1110.3193L, 2022A&A...662A.112E}, Rubin Observatory \citep{2022arXiv220307220B}, and Roman Space Telescope \citep{2015arXiv150303757S}.

In this work, we conduct a pilot study for CSST to validate the reliability of the algorithm developed by \cite{zhouAutomaticMorphologicalClassification2022} and \cite{fangAutomaticClassificationGalaxy2023} with COSMOS I-band images since these images are high-resolution and similar to the imaging data of the CSST. In brief, we perform our two-step algorithm to galaxies with $I_{\rm mag} < 25$ at redshifts $0.2<z<1.2$ in the COSMOS field to get their morphological types. Additionally, both parametric and nonparametric morphologies for these galaxies are also estimated to investigate the consistency between our classification results and these morphological parameters. The results confirm that our classification results align with the expected relationships with the morphological parameters, underscoring the reliability of our algorithm.

The paper's structure is as follows: In Section \ref{sec:data}, an overview of the COSMOS program and our sample selection criteria are provided. Section \ref{sec:method} briefly introduces the methodology employed for galaxy morphological classification. Our classification results are presented in Section \ref{sec:results}, followed by the conclusion in Section \ref{sec:summary}. Throughout this work, we adopt a flat $\rm \Lambda$ cold dark matter ($\rm \Lambda$CDM) cosmology with $H_0 = 70\ \rm km\ s^{-1}\ Mpc^{-1}$, $\Omega_{\rm m} = 0.3$, and $\Omega_{\rm \Lambda} = 0.7$, along with a \cite{chabrierGalacticStellarSubstellar2003} initial mass function. 

\section{DATA SET AND SAMPLE SELECTION} \label{sec:data}
\subsection{COSMOS} \label{subsec:COSMOS}
The COSMOS field was strategically designed to explore the intricate connections between galaxy evolution, star formation, active galactic nuclei, dark matter, and large-scale structure within the redshift range $0.5<z<6$. This comprehensive survey spans a wide range of wavelengths, from X-ray to radio, and covers an area of approximately $\rm 2\ deg^2$. In this study, the high-resolution images captured by HST/ACS in the F814W filter are capitalized, encompassing an area of roughly 1.64 $\rm deg^2$ within the COSMOS field. This dataset represents the largest continuous field observed by HST ACS as yet. The original images comprise approximately 590 pointings, with an average exposure time of 2028 seconds per pointing. These images were meticulously processed by \cite{koekemoerCOSMOSSurveyHubble2007} using the STSDAS {\tt\string Multidrizzle} package \citep{2003hstc.conf..337K}, resulting in images with a pixel scale of $0\farcs03$ and a 5$\sigma$ depth of 27.2 AB mag within a $0\farcs24$ diameter aperture. In the subsequent analysis, all morphological measurements are based on these high-resolution HST I-band images.

\subsection{COSMOS2020 Catalogue} \label{subsec:catalogue}

The sample utilized in this study is constructed based on the ``Farmer'' COSMOS2020 catalog \citep{weaverCOSMOS2020PanchromaticView2022}, which provides comprehensive photometric information spanning 35 bands from ultraviolet to near-infrared. \cite{weaverCOSMOS2020PanchromaticView2022} further estimated some physical properties of galaxies in this field, including photometric redshift and stellar mass, through spectral energy distribution (SED) fitting with galaxy models, using this enhanced photometric dataset.

Redshifts were determined employing two distinct codes, {\tt\string EAZY} \citep{brammerEAZYFastPublic2008}  and {\tt\string LePhare} \citep{ilbertAccuratePhotometricRedshifts2006}. For our analysis, we adopt the {\tt\string LePhare}-derived redshifts, as Figure 15 of \cite{weaverCOSMOS2020PanchromaticView2022} illustrated their superior reliability within the magnitude range under consideration. Redshift estimation involves a library of 33 galaxy templates sourced from \cite{bruzualStellarPopulationSynthesis2003} and \cite{ilbertCOSMOSPHOTOMETRICREDSHIFTS2009}. Additionally, various dust extinction/attenuation curves are employed, encompassing the starburst attenuation curve introduced by \cite{calzettiDustContentOpacity2000}, the SMC extinction curve from \cite{1984A&A...132..389P}, and two variations of the Calzetti law, including the 2175 Å bump. Then the photometric redshift ($z_{\rm LePh}$) is defined as the median value derived from the redshift likelihood function.

Subsequently, with redshift fixed to $z_{\rm LePh}$, the {\tt\string LePhare} fitting code was executed once more to extract stellar mass and other pertinent physical properties. This stage considered \cite{bruzualStellarPopulationSynthesis2003} stellar populations and a range of stellar formation histories (SFHs), including exponentially declining SFHs and delayed $\tau$ SFHs. For additional in-depth details, please refer to the works of \cite{weaverCOSMOS2020PanchromaticView2022} and \cite{laigleCOSMOS2015CATALOGEXPLORING2016}.

\subsection{Sample Selection} \label{subsec:sample}

\begin{figure}[htb!]
\centering
\includegraphics[width=0.48\textwidth, height=0.384\textwidth]{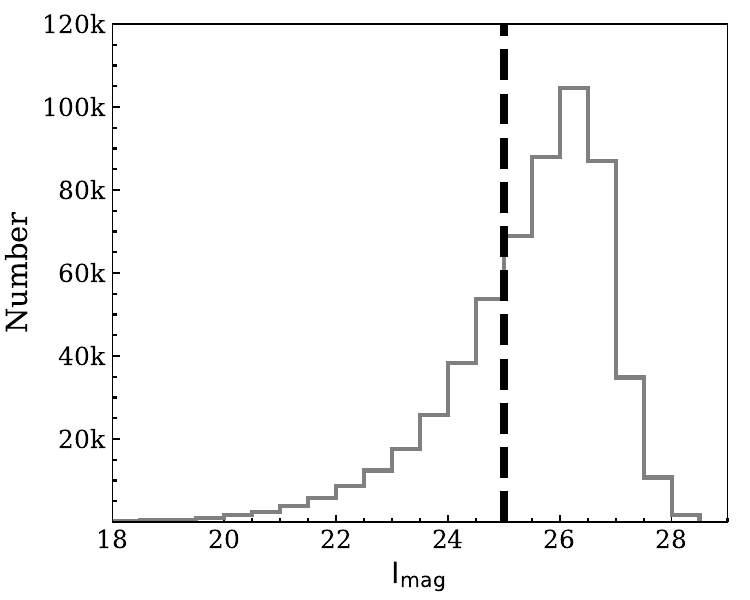}
\caption{The distribution of I band magnitudes in the COSMOS field. A magnitude limit of $I_{\rm mag}<25$ is imposed.}
\label{fig1}
\end{figure}

In this work, we have selected galaxies from the COSMOS2020 Catalog based on the following criteria: (1) $I_{\rm mag}$ $<$ 25 mag, excluding galaxies that are too faint to obtain reliable morphological measurements; 
(2) $0.2 < z < 1.2$, ensuring that the morphology is estimated in the rest-frame optical band; (3) $\rm FLAG_{ COMBINE}=0$, which means flux measurements are not influenced by bright stars and the objects are not on the edges of images, guaranteeing reliable photometric redshift and stellar mass estimations; (4) signal to noise ration larger than 5 ($\rm S/N > 5$), ensuring a real detection. Moreover, sources with bad pixels have also been excluded. As a result, the final sample comprises a total of 99,806 galaxies. The distribution of I-band magnitudes in the COSMOS field, along with the magnitude selection threshold of 25, is illustrated in Figure \ref{fig1}.

\begin{figure}[htb!]
\centering
\includegraphics[width=0.48\textwidth, height=0.35\textwidth]{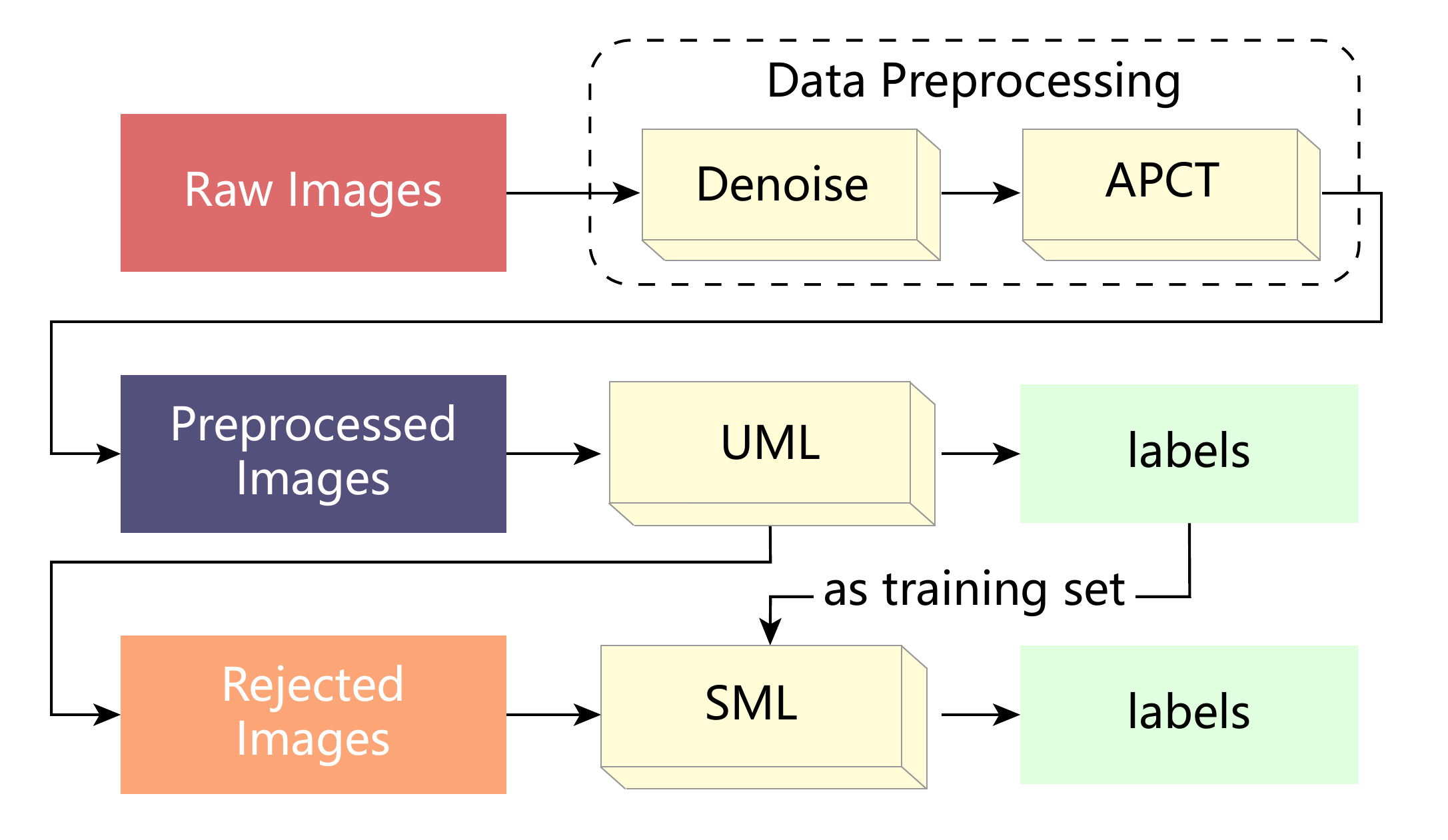}
\caption{Framework of the morphological classification model employed in this work.}
\label{fig2}
\end{figure}

\section{Method for Morphological Classification} \label{sec:method}

In this section, an overview of our {\tt\string Usmorph} algorithm employed for galaxy morphological classification is provided (as shown in Figure \ref{fig2}). This comprises three sections(data preprocessing, UML method, and SML method), and the detailed content is as follows.

\subsection{Data Preprocessing} \label{subsec:preprocessing}

\begin{figure*}[htb!]
\centering
\includegraphics[width=0.9\textwidth, height=0.5\textwidth]{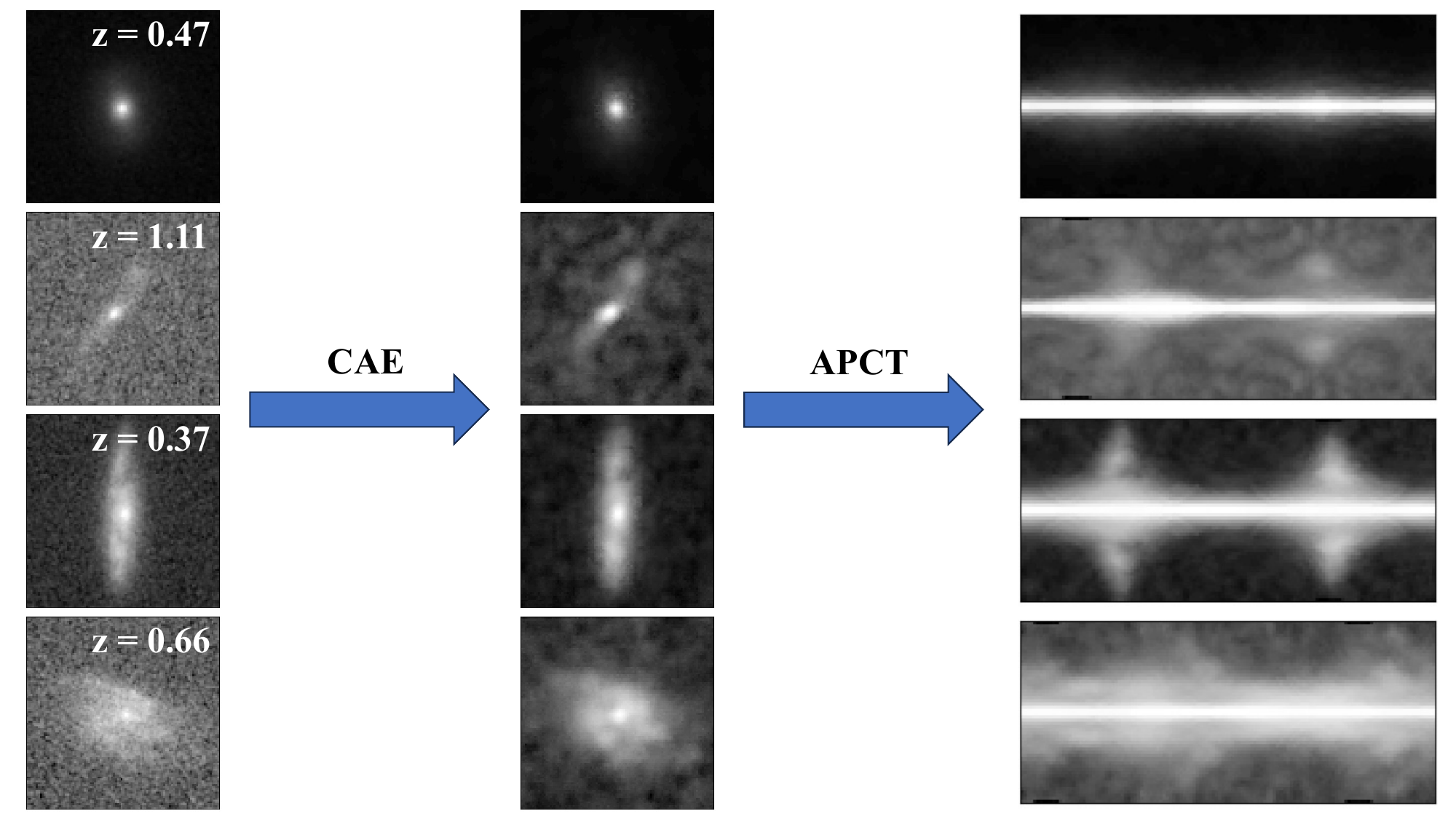}
\caption{Some examples of image preprocessing, the left column displays original images of four randomly selected galaxies, the middle column shows the corresponding denoised images, and the right column displays images that have undergone a polar coordinate transformation. The redshift information of these chosen galaxies is shown in the first column.}
\label{fig3}
\end{figure*}

In our previous work, \cite{zhouAutomaticMorphologicalClassification2022} employed a 28$\times$28-pixel cutout for each galaxy within the CANDELS field. However, given that the pixel scale of the COSMOS I-band images is $0\farcs03$ per pixel, which is notably finer than the $0\farcs06$ pixel scale in the CANDELS field, and that the redshift range considered in this study is quite lower than that of \cite{zhouAutomaticMorphologicalClassification2022}, larger 100 $\times$ 100-pixel cutouts are generated for galaxies in our sample, with all galaxies centered in each cutout. To ensure that this cutout is sufficiently large to encapsulate adequate information for the majority of selected galaxies, we have checked that the half-light radius of approximately 97\% of the galaxies in our sample is less than 0.12 arcsec (equivalent to 40 pixels). 
To prevent an excessively large cutout size, we have also checked and found that approximately 11\% of galaxies have a half-light radius greater than 0.09 arcsec (30 pixels). To balance the training time and ensure most of the source information, we think this size is reasonable when considering that most cutouts (96\%) contain only one galaxy.
Moreover, before the following steps, a max-min normalization pretreatment is also applied to each cutout, following the methods described in \cite{zhouAutomaticMorphologicalClassification2022}.


We update the UML method developed in \cite{zhouAutomaticMorphologicalClassification2022} by introducing some data preprocessing steps, enhancing the reliability of our results. Previously, some studies had shown that the distribution of image S/N can affect the performance of machine learning since noise can disturb image features and lead to misclassifications \citep{liuNetworksPixelsEmbedding2022}. Fortunately, \cite{10.1007/978-3-319-75193-1_50} had demonstrated that this problem can be overcome with noise reduction. 
CAE is demonstrated to be highly effective in noise reduction via automatically extracting image features and reconstructing images with the extracted features (e.g., \citealt{masciStackedConvolutionalAutoEncoders2011, fangAutomaticClassificationGalaxy2023, daiClassificationGalaxyMorphology2023a}). 
In the CAE process, operations of convolution and pooling encode the pixels and give encoded features with a lower dimension. Then, denoised results can be obtained by applying deconvolution and unpooling to these features. To make our classification results more reliable, we also perform noise reduction with CAE to enhance the image quality of our samples. In our CAE algorithm, we consider 2 layers during the encoding process, each layer consisting of convolution and max-pooling operations. Then, through a fully connected layer, we represent the features of galaxies as a 40-dimensional vector. The decoding process is the inverse of the encoding process, and the configurations used during encoding and decoding are identical.  An schematic diagram of this process can be found in Figure 2 of \cite{zhouAutomaticMorphologicalClassification2022}. We build our algorithm based on {\tt\string TensorFlow}\footnote{\url{https://www.tensorflow.org}}\citep{2016arXiv160304467A}. During the convolution operations, the channel size is set to 16 with a kernel size of $5\times 5$. Subsequently, in the max-pooling process, a $2\times 2$ pixel block is consolidated into a single pixel. Due to memory constraints, the batch size is set to 8. The activation function is the Rectified Linear Unit function \citep{2018arXiv180308375A}, while the loss function is the Mean Square Error \citep{lehmann2006theory}. We also employ a learning rate scheduler (Exponential Decay, \citealt{2019arXiv191007454L}) to dynamically adjust our learning rate, starting with an initial value of $3\times 10^{-4}$. We do not use batch normalisation, or dropout, or regularisation in this work. We train our program for 32 epochs, extending the training by approximately 10 additional epochs beyond the point where the validation loss plateaued to get a better denoising effect.
Figure \ref{fig3} provides a visual comparison before and after noise reduction: in the left column, raw images of randomly selected galaxies from our sample are presented, and in the middle column, the results after noise reduction are showcased. It is evident that the image quality is substantially improved after noise reduction while retaining the essential morphological features. 

Furthermore, a number of groups point out that the standard CNN models have poor robustness to rotations of images, which may lead to misclassification of the galaxy morphological type after rotation (e.g., \citealt{8462105, YaoFCHG19}). To overcome this shortcoming, \cite{fangAutomaticClassificationGalaxy2023} proposed an adaptive polar coordinate transformation (APCT) method in the data preprocessing procedure. Compared to traditional data augmentation and conventional polar coordinate unwrapping, \cite{fangAutomaticClassificationGalaxy2023} found that APCT can significantly improve the accuracy of CNNs when images are rotated. Moreover, by ignoring information about image orientation, APCT can make models focus on other more important features of galaxies, which makes it an important preprocessing method. Here we provide a brief introduction to this method (see \citealt{fangAutomaticClassificationGalaxy2023} for more details).
Essentially, the pixels with the maximum and minimum flux values in the images are identified as the brightest and darkest points, respectively. Subsequently, the line connecting the brightest and darkest points is designated as the polar axis. Then, the axis is rotated counterclockwise by an increment of 0.05 rad each time. For each discrete rotation, the axis traverses numerous pixels of the raw images. By stacking the pixels along this rotating axis while rotating, the entire image can be
unfolded into polar coordinates. Finally, the images are mirrored to highlight morphological features since CNN is more sensitive to information at the center of the image. To make our results rotationally invariant, We apply this method to our samples. In the right column of Figure \ref{fig3}, we present the corresponding results after implementing APCT.

\subsection{UML Process} \label{subsec:uml}

\begin{figure*}
\centering
\includegraphics[width=1.0\textwidth, height=0.55\textwidth]{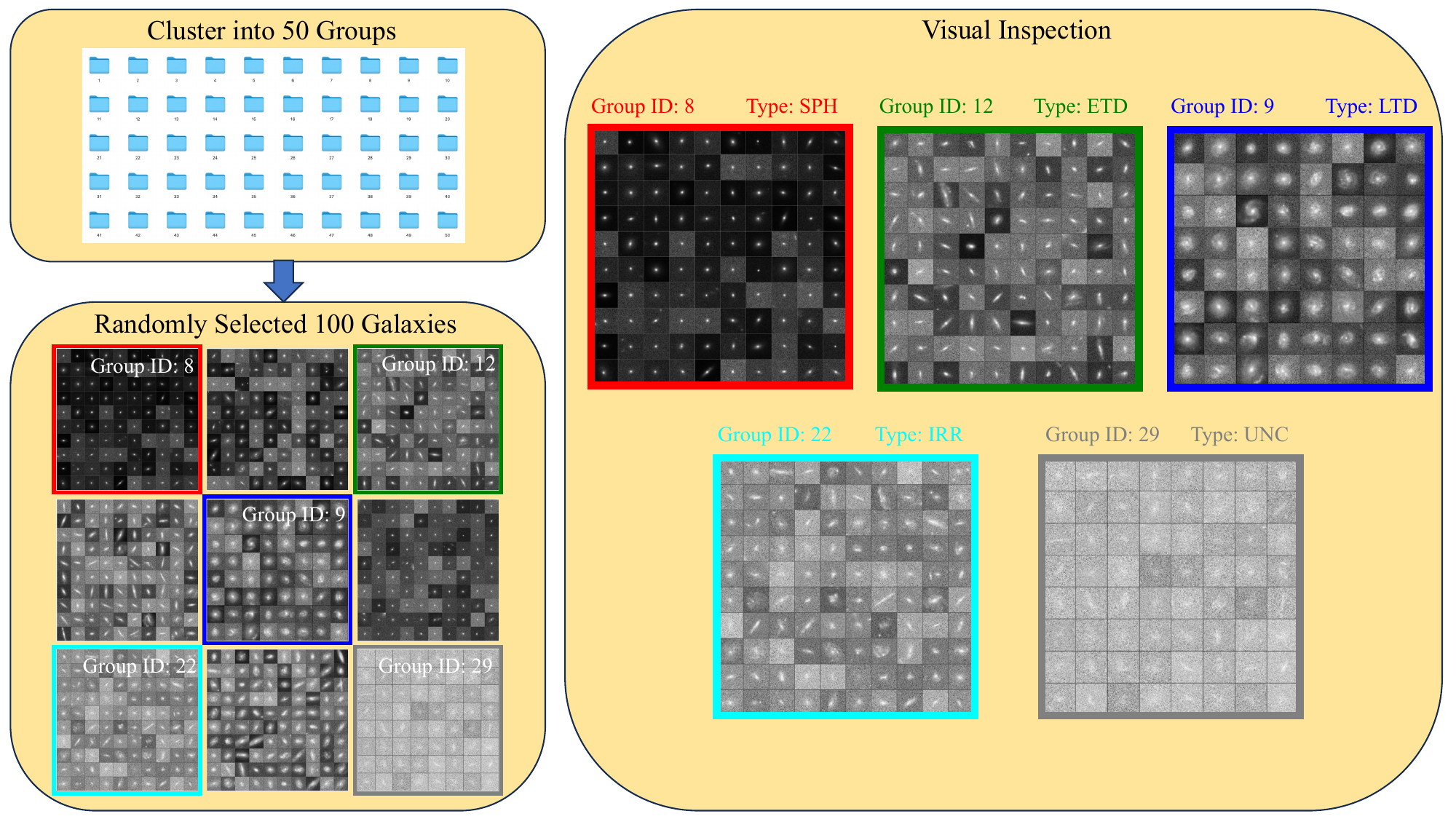}
\caption{Illustration of the post-hoc visual inspection for label alignment. 100 galaxies are randomly selected from each group (only 64 galaxies are selected for Group 9 because the total number of galaxies in this group is less than 100) for visual inspection. Since galaxies within each group share similar features, we are able to quickly categorize each group into five types (SPH, ETD, LTD, IRR, and UNC) by simultaneously assessing the overall morphological features of these selected galaxies.}
\label{fig4}
\end{figure*}

After completing the data preprocessing steps, we apply the UML method, as outlined in \cite{zhouAutomaticMorphologicalClassification2022}, to obtain classification results for some of our samples. The UML method comprises two main steps: (1) Using the images processed with noise reduction and APCT, relevant features are extracted with CAE\footnote{In the preprocessing part, we use CAE to perform noise reduction. In this step, we run CAE once more to extract features.The configuration of CAE used here for feature extraction is the same as the one used for noise reduction}, then these extracted 40-dim features are used for the next step; (2) a bagging-based multi-clustering approach is employed to cluster galaxies with similar features into distinct groups. In this second step, three different clustering models are employed simultaneously: the k-means clustering algorithm \citep{hartigan1979algorithm}, the agglomerative clustering algorithm \citep{murtaghSurveyRecentAdvances1983, murtaghWardHierarchicalAgglomerative2014}, and the Balanced Iterative Reducing and Clustering using Hierarchies (BIRCH) algorithm \citep{zhangBIRCHEfficientData1996}. The sample is clustered into 50 categories by each model separately. By setting the K-means labels as the primary labels, we assign labels to the groups of other models according to the highest frequency of the K-means label in that group. Then, the ``majority wins'' strategy is used in voting. The sources that the three models reach a consensus in voting are retained, while those that do not reach a consensus are discarded. This voting strategy can improve the clustering quality, thereby leading to a more reliable classification, as demonstrated in \cite{zhouAutomaticMorphologicalClassification2022} and \cite{2023arXiv231108995L}.

After removing galaxies with inconsistent voting results, 36,604 galaxies are neatly grouped into 50 groups. The morphological type of these 50 groups is determined through post-hoc visual inspection for label alignment (refer to \citealt{2023arXiv231108995L} for further details). Specifically, owing to the high purity of the clustering, galaxies within each group exhibit similar features. Consequently, the morphological type for a particular group can be obtained by visually inspecting only a subset of galaxies from that group. In this study, we randomly select 100 galaxies from each group for visual inspection. By evaluating the overall morphological features of these randomly selected 100 samples simultaneously, we categorize each group into one of five types: SPH, ETD, LTD, IRR, and UNC. Finally, these 36,604 galaxies are classified into five distinct categories: SPHs (8233), ETDs (5322), LTDs (6320), IRRs (9468), and UNCs (7261). Figure \ref{fig4} illustrates an example of this process. 

\subsection{SML Process} \label{subsec:sml}

\begin{figure}
\centering
\includegraphics[width=0.42\textwidth, height=0.3\textwidth]{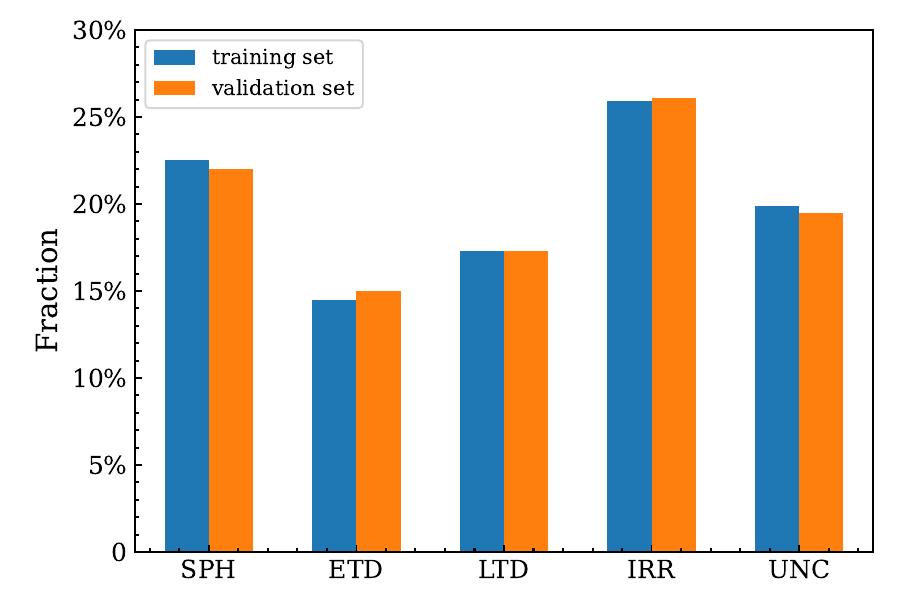}
\caption{The distribution of galaxy morphological types in the training and validation sets.}
\label{fig5}
\end{figure}

\begin{figure*}
\centering
\includegraphics[width=1.0\textwidth, height=0.3\textwidth]{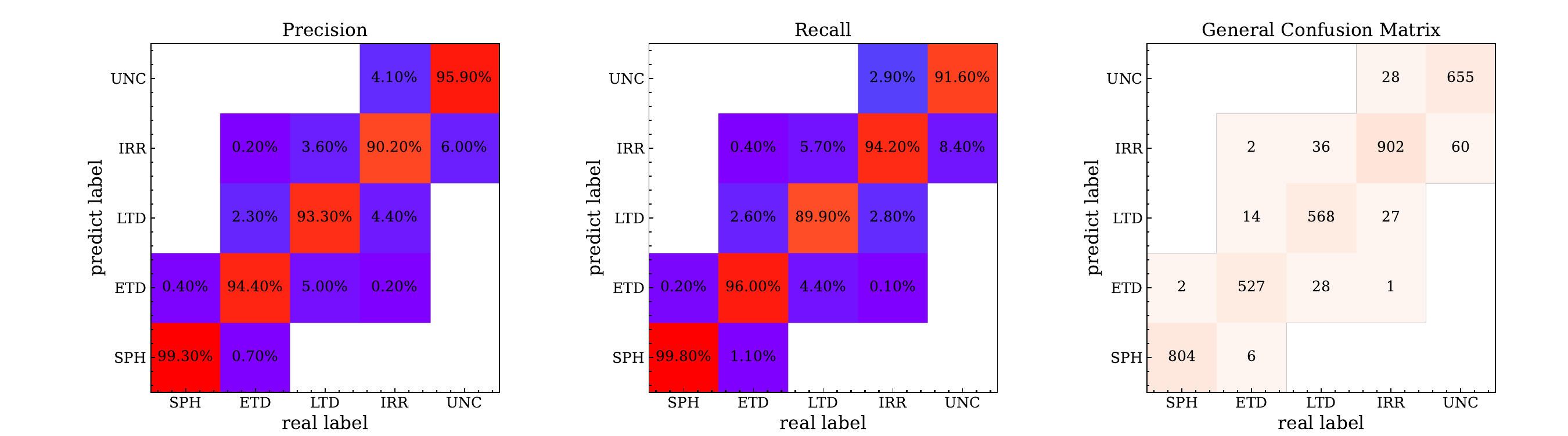}
\caption{The left and middle panels represent the precision and recall of the GoogleNet model, with both overall precision and recall exceeding 94\%, indicate that the GoogleNet model can effectively distinguish between different types of galaxies. The right panel represents the general confusion matrix.}
\label{fig6}
\end{figure*}

By excluding 63,202 sources with inconsistent voting results, a set of 36,604 galaxies with reliable morphological labels are obtained through our UML clustering process. Then these 36,604 well-classified sources are used as a training set to conduct SML for the remaining 63,202 galaxies. Based on the findings of \cite{fangAutomaticClassificationGalaxy2023}, who demonstrated that GoogLeNet performs well in classifying deep-field galaxies, we employ the GoogLeNet algorithm \citep{szegedyGoingDeeperConvolutions2015} as a supervised classification model. 

To prevent overfitting, the labeled galaxies are randomly divided into a 9:1 ratio of training set (32,944) and validation set (3660), following the same proportion as in \cite{fangAutomaticClassificationGalaxy2023}.
To ensure the robustness of our GoogleNet model, we also examined the distribution of morphological types in the training and validation sets. The results are shown in Figure \ref{fig5}, from which it can be clearly seen that these two sets have the same distribution of morphological types. Additionally, we have also examined the distribution of the training set and validation set in some other physical parameter (including $M_{\ast}$, redshift, and some other morphological parameters) space and find that they alao exhibit similar distributions. Figure \ref{fig6} displays the precision and recall rates of the GoogLeNet model, which is estimated based on the validation set. It is evident that the overall accuracy rate is approximately 94\%, indicating that GoogLeNet performs well in classifying galaxies of all types.  With this SML method, we obtain morphological classifications for the remaining 63,202 galaxies.

\section{Results and Discussion} \label{sec:results}

\subsection{Overall morphological classification results} \label{subsec:class_result}

\begin{figure*}[htb!]
\centering
\includegraphics[width=0.95\textwidth, height=0.57\textwidth]{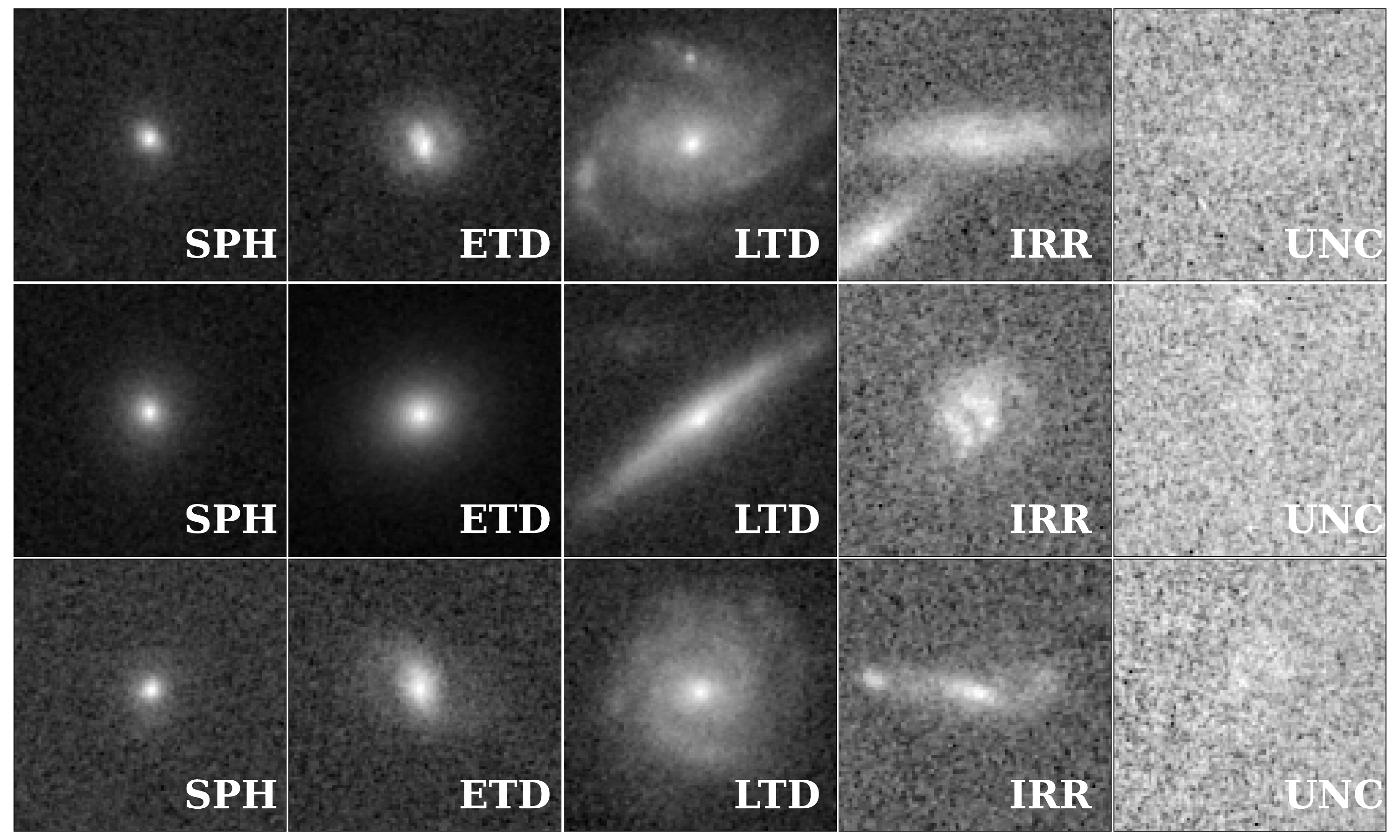}
\caption{Some examples of the final classification results. From left to right, they are SPHs, ETDs, LTDs, IRRs, and UNCs.}
\label{fig7}
\end{figure*}

\begin{figure}[htb!]
\centering
\includegraphics[width=0.48\textwidth, height=0.4\textwidth]{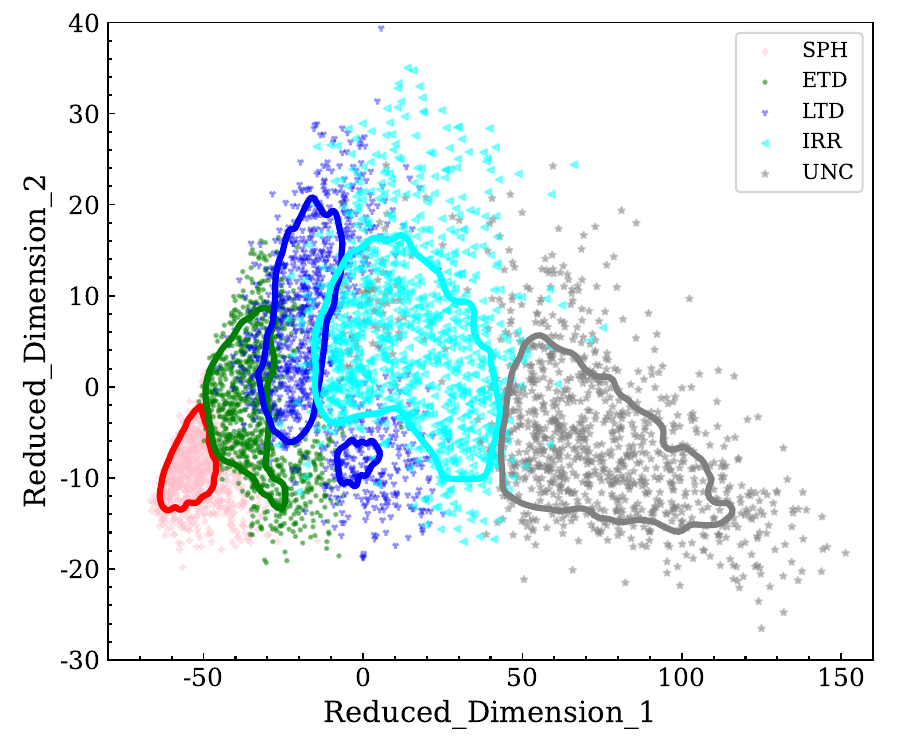}
\caption{The t-SNE diagram of randomly selected 5000 galaxies from the final classification result, where red, green, blue, cyan, and grey represent SPHs, ETDs, LTDs, IRRs, and UNCs, respectively. Additionally, unless otherwise stated, the same color scheme is adopted in the following analyses. The solid lines of different colors in the figure represent the corresponding 1 sigma contours. Different types of galaxies exhibit distinct boundaries on the t-SNE diagram, indicating that our classification algorithm can effectively distinguish between different types of galaxies.}
\label{fig8}
\end{figure}

\begin{deluxetable}{ccccccc}
\centerwidetable
\tablecaption{The number of galaxies classified into different types \label{tab1}}
\tablehead{\colhead{TYPE} & \colhead{SPH} & \colhead{ETD} & \colhead{LTD} & \colhead{IRR} & \colhead{UNC}
& \colhead{TOTAL}} 
\startdata  
UML &  8233 & 5322 & 6320 & 9468 & 7261 & 36,604\\
SML & 6967 & 12,047 & 14,823 & 19,497 & 9868 & 63,202\\
TOTAL & 15,200 & 17,369 & 21,143 & 28,965 & 17,129 & 99,806\\
\enddata
\end{deluxetable}

In preparation for future large-scale surveys, there is a growing demand for efficiently obtaining morphological information for a significant number of galaxies within a short time frame. In this study, we have successfully classified the morphologies of approximately 100,000 galaxies in the COSMOS field using I-band images, which include 15,200 SPHs, 17,369 ETDs, 21,143 LTDs, 28,965 IRRs, and 17,129 UNCs. Detailed classification results are shown in Table \ref{tab1}. The whole task (both UML and SML) is completed within less than one day, which means this method has potential for application in future large-field sky surveys.

To test our classification, we conduct visual inspections on a randomly sampled subset of samples from the final results. The visual examination confirms that SPHs tend to be compact and bulge-dominated, whereas LTDs exhibit extended structures with disk dominance. Some LTDs also display prominent spiral arm structures. ETDs are characterized by their relatively less compact form with a bright nuclear region and a disk component. IRRs encompass galaxies with irregular structures or merger signatures. UNCs are typically galaxies that cannot be confidently identified due to poor S/Ns. Some examples of the classification results are presented in Figure \ref{fig7}, which demonstrates the effectiveness of our algorithm in distinguishing galaxies with different morphological types.

Furthermore, the t-distributed Stochastic Neighbor Embedding (t-SNE) technique is adopted to assess the classification results, which maps high-dimensional data into a two- or three-dimensional space, making it suitable for visualization and inspection \citep{JMLR:v9:vandermaaten08a}. To enhance clarity, 5000 randomly selected galaxies are presented using the t-SNE technique in Figure \ref{fig8}, which is based on the extracted 40-dim features in the UML process. Considering that the axes do not have corresponding physical meanings, here we simply use ``Reduced\_Dimension\_1'' and ``Reduced\_Dimension\_2'' to represent the axis labels. The solid lines of different colors in the figure represent the corresponding 1 sigma contours. This figure reveals distinct boundaries between different galaxy types, indicating that each class of galaxies in our classification indeed possesses distinct characteristics. We note that LTDs appear to be divided into two parts in the t-SNE diagram, which may be attributed to projection effects. There is also overlaps between different groups. This may also be caused by the morphological similarities between different groups, which is expected by galaxy evolution scenario. Since the t-SNE diagram can only provide a qualitative overview of the results, we will further validate our classification results in the following sections.

\subsection{Test of Morphological Parameters} \label{subsec:morph}
The structural parameters of galaxies are closely related to their morphological types. For example, it is commonly accepted that elliptical galaxies tend to have S\'{e}rsic index greater than 2, while disk galaxies typically exhibit S\'{e}rsic index less than 2 (e.g., \citealt{fisherSTRUCTURECLASSICALBULGES2008, blantonPhysicalPropertiesEnvironments2009a}). Nonparametric structural parameters also offer valuable insights into galaxy morphological types, as different galaxy categories often occupy distinct positions in parameter space (e.g., \citealt{lotzEvolutionGalaxyMergers2008, yaoEvolutionNonparametricMorphology2023}).

In this section, we investigate the classification results using galaxy morphological parameters. Since extensive researches have already explored the correlations between morphologies and other morphological parameters (e.g., \citealt{guMorphologicalEvolutionAGN2018, zhouAutomaticMorphologicalClassification2022, daiClassificationGalaxyMorphology2023a}), our analysis focuses exclusively on massive galaxies with stellar masses exceeding $10^{9}M_{\odot}$. We exclude the UNCs subclass from our analysis due to the difficulty in measuring morphological parameters for these galaxies, primarily stemming from their relatively low S/Ns.

\subsubsection{Parametric Measurements} \label{subsec:parametric}

\begin{figure*}[htb!]
\centering
\includegraphics[width=0.95\textwidth, height=0.38\textwidth]{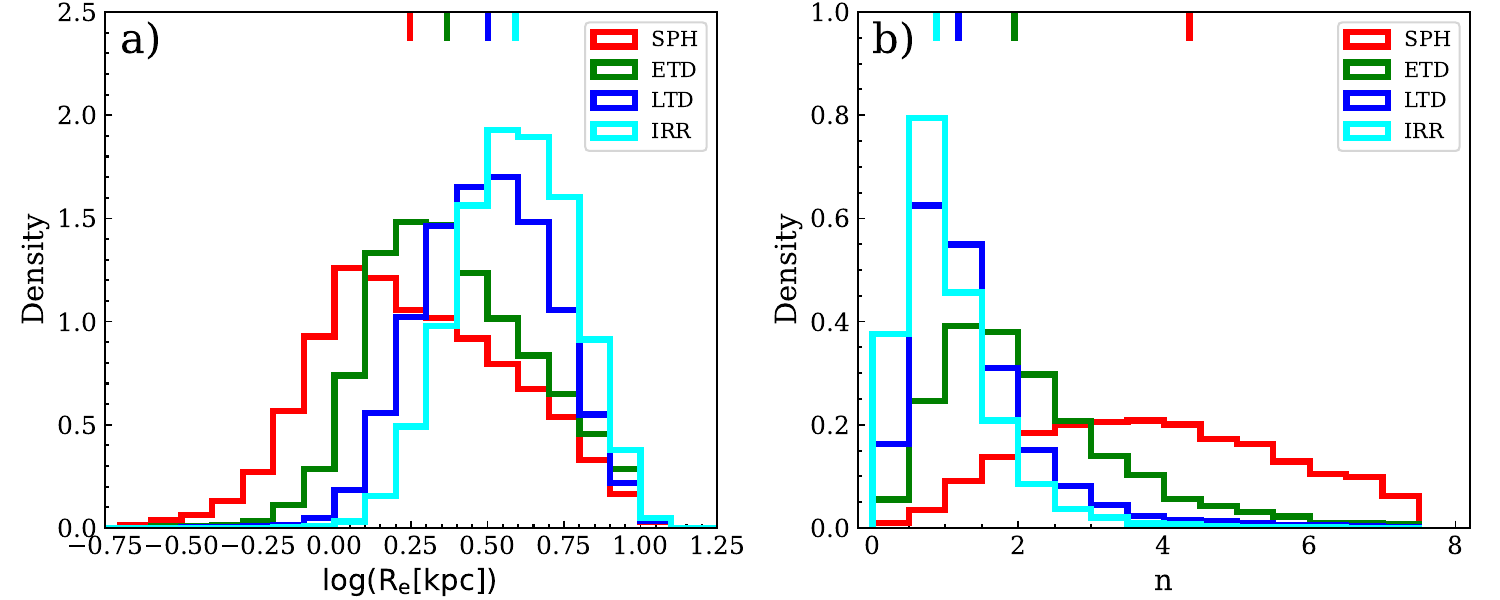}
\caption{The distribution of effective radius (left panel) and S\'{e}rsic index (right panel) for different types of galaxies. The bars at the top represent the median values of $\log(r_e)$ and n for different galaxy types. It is evident that from SPHs to IRRs, the effective radius of galaxies gradually increases, while the S\'{e}rsic index gradually decreases.}
\label{fig9}
\end{figure*}

In this study, the measurements of parametric morphology for galaxies in the COSMOS field are conducted using the {\tt\string GALAPAGOS} software \citep{bardenGalapagosPixelsParameters2012a, hausslerGalapagos2GalfitmGAMA2022a}, which serves as a wrapper for both {\tt\string SExtractor} \citep{bertinSExtractorSoftwareSource1996} and {\tt\string GALFIT} \citep{pengDetailedStructuralDecomposition2002a}. The software utilizes a single S\'{e}rsic model to fit the surface brightness profile of each galaxy and measure their S\'{e}rsic index and effective radius ($r_e$).

To ensure the reliability of our analysis and eliminate the influence of galaxies that cannot be fully covered by our $100\times 100$ pixel cutout, we excluded galaxies for which the estimated $r_e$ exceeded 40 pixels. This removal affects less than 5\% of our galaxy sample, demonstrating that it had no significant impact on our results.

The left panel of Figure \ref{fig9} illustrates the distribution of effective radius for four galaxy subclasses: SPHs, ETDs, LTDs, and IRRs. The median effective radius for these subclasses are 
1.77, 2.33, 3.18, and 3.90 kpc, respectively. Notably, SPHs tend to have smaller sizes compared to the other subclasses, and the effective radius increases in the order from SPHs to ETDs, LTDs, and IRRs.
The right panel of Figure \ref{fig9} presents the distributions of S\'{e}rsic indices for the same four galaxy subclasses. SPHs exhibit a higher degree of compactness with a median S\'{e}rsic index of 4.34, whereas ETDs, LTDs, and IRRs have median S\'{e}rsic index of 1.94, 1.18, and 0.88, respectively. In summary, the distributions of S\'{e}rsic index and effective radius for different galaxy types align with our expectations regarding the relationships between galaxy types and structural parameters. 

\subsubsection{Nonparametric Measurements} \label{subsubsec:nonparametric}

\begin{figure*}[htb!]
\centering
\includegraphics[width=0.95\textwidth, height=0.432\textwidth]{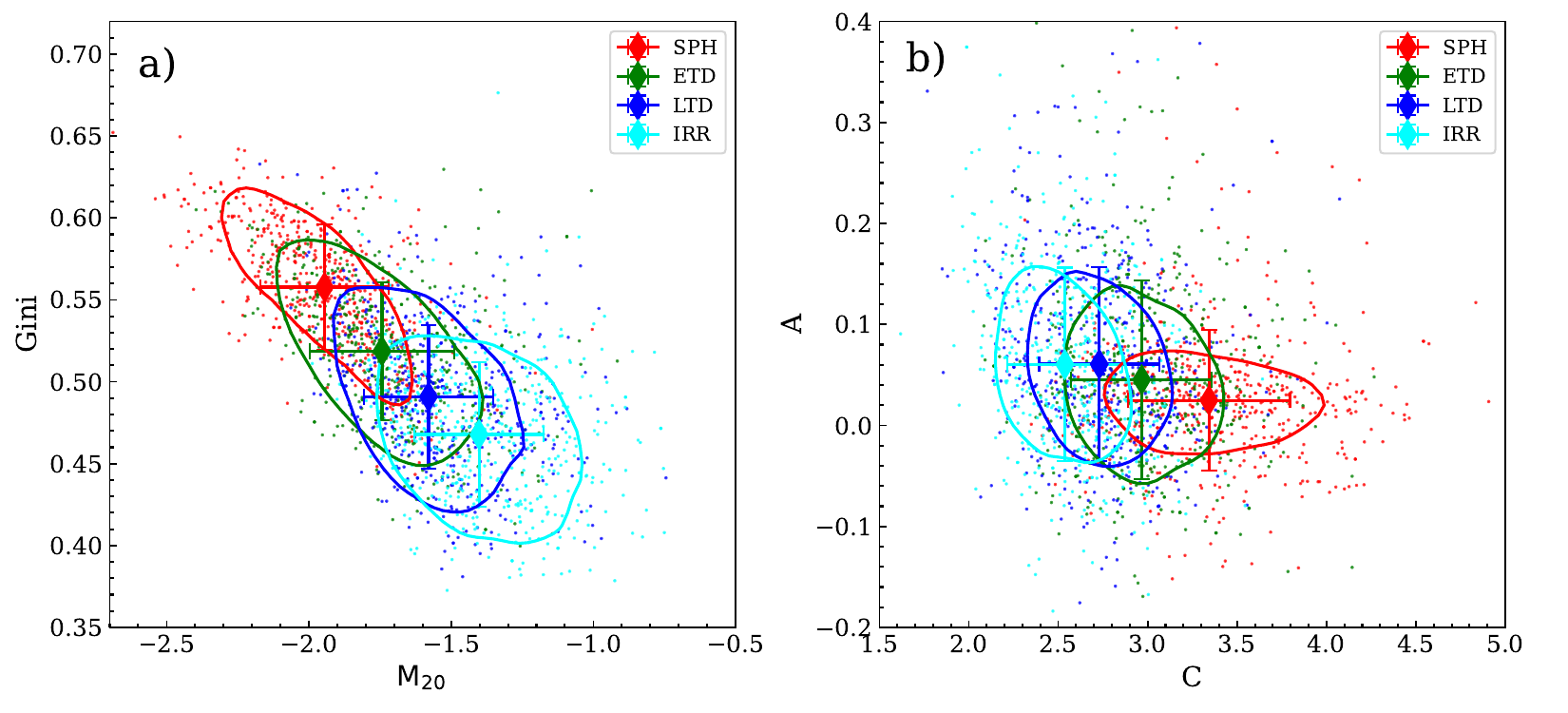}
\caption{The distribution of four subclasses in the Gini-$M_{\rm 20}$ (left) and C-A (right) space. The contour represents 50\% of the specified galaxies. The diamonds with different colors correspond to the median values of the parameters and the error bars represent the corresponding standard deviations. }
\label{fig10}
\end{figure*}

\begin{figure*}[htb!]
\centering
\includegraphics[width=0.95\textwidth, height=0.38\textwidth]{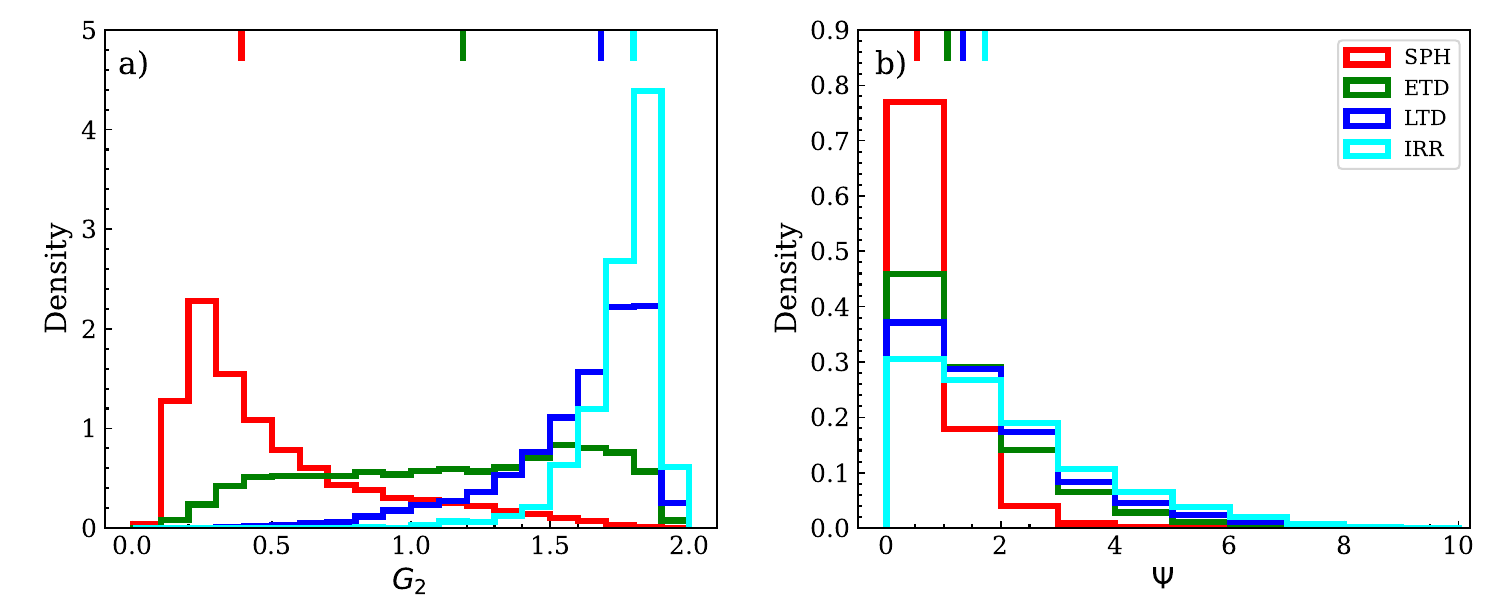}
\caption{The distribution of $G_2$ (left panel) and $\Psi$ (right panel) for different types of massive galaxies. The bars at the top represent the median values of the corresponding parameters.}
\label{fig11}
\end{figure*}

\begin{figure*}[htb!]
\centering
\includegraphics[width=0.95\textwidth, height=0.28\textwidth]{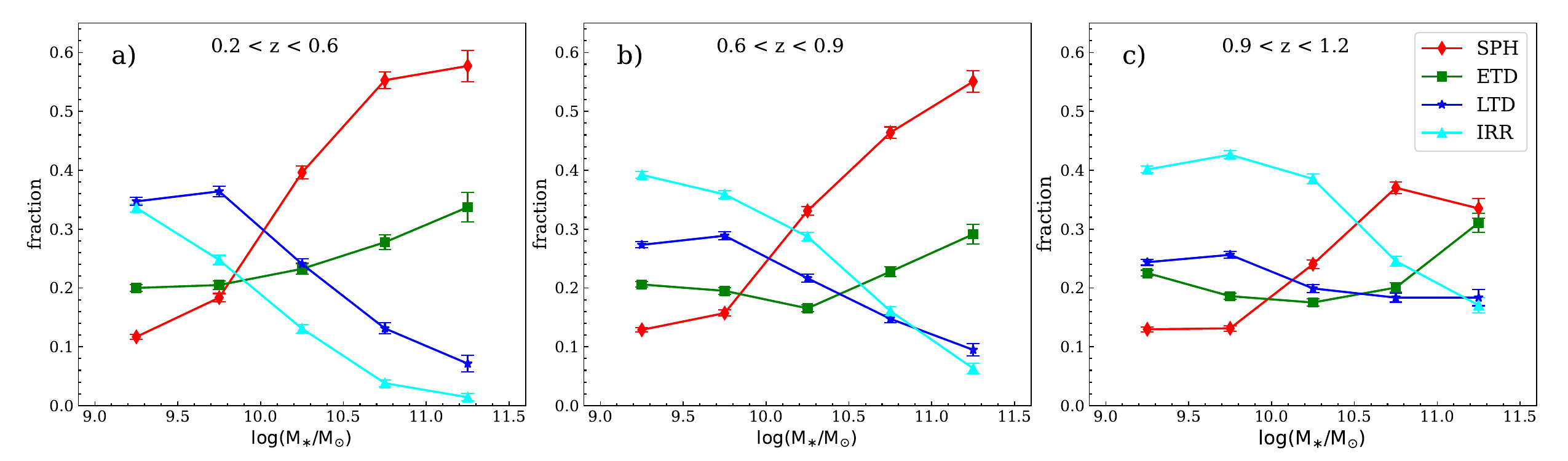}
\caption{The proportion of different types of galaxies as a function of stellar mass, with different panels displaying the results at different redshifts bins (panel a: $0.2<z<0.6$; panel b: $0.6<z<0.9$; panel c: $0.9<z<1.2$).}
\label{fig12}
\end{figure*}

Since the morphological structures of high-redshift galaxies are likely to be irregular, it can be challenging to fit the brightness distribution of these high-redshift galaxies with empirical functions. Therefore, some nonparametric morphological parameters are developed to describe various features of galaxies. The most widely used parameters are CAS statistics. C (concentration) is developed to quantify the concentration of light within a galaxy's central region compared to its outer regions (e.g., \citealt{2000AJ....119.2645B, conseliceRelationshipStellarLight2003}). A (asymmetry) could characterize the degree of asymmetry exhibited by a galaxy after a $180^{\circ}$ rotation (e.g., \citealt{conseliceAsymmetryGalaxiesPhysical2000, conseliceRelationshipStellarLight2003}). S (clumpiness) represents the proportion of light in a galaxy that is distributed in clump patterns. Besides the CAS system, the Gini$-M_{\rm 20}$ system has also been adopted in many works. Gini coefficient represents the distribution of light within a galaxy and a higher value indicates a more unequal distribution (e.g., \citealt{lotzNewNonparametricApproach2004, lotzEvolutionGalaxyMergers2008}). $M_{\rm 20}$ (the normalized second-order moment of the brightest 20\% of the galaxy’s flux) indicates whether light is concentrated in a galaxy, and a more concentrated result (a more negative $M_{\rm 20}$) does not imply that light is central concentration, instead, the light may be concentrated in any location within the image (e.g., \citealt{lotzNewNonparametricApproach2004, lotzEvolutionGalaxyMergers2008}). More recently, a new suit of parameters was introduced by \cite{freemanNewImageStatistics2013}, that includes M (multimode, which quantifies the area ratio between two most ``prominent'' clumps within a galaxy), I (intensity, which represents the light ratio between the two brightest subregions within a galaxy.), and D (Deviation, which represents the distance between the image light-weighted center and the brightest peak.)


\cite{rodriguez-gomezOpticalMorphologiesGalaxies2019} had developed a python package {\tt\string statmorph} for calculating the nonparametric morphology of galaxies, including Gini, $M_{\rm20}$, C, A, S, M, I, and D. Based on this package, \cite{yaoEvolutionNonparametricMorphology2023} had optimized the code with {\tt\string Cython} \citep{behnelCythonBestBoth2011} and increased the calculation speed by one order of magnitude, which is suitable for the large-scale survey. In addition, \cite{yaoEvolutionNonparametricMorphology2023} had also incorporated some additional optional parameters into {\tt\string statmorph}, including the color dispersion ($\xi$; \citealt{papovichInternalUltravioletOptical2003}), Multiplicity ($\Psi$; \citealt{lawPhysicalNatureRest2007}), and the second gradient moment ($G_2$; \citealt{rosaGradientPatternAnalysis2018}). The improved code has been named {\tt\string statmorph\_csst}. The reliability of {\tt\string statmorph} has been extensively validated in previous studies. \cite{yaoEvolutionNonparametricMorphology2023} conducted tests to compare measurement results between {\tt\string statmorph\_csst} and the original {\tt\string statmorph}. The results, presented in their Appendix 1, indicate that the optimization has virtually no impact on the measured results of morphological parameters, underscoring the reliability of {\tt\string statmorph\_csst}. In this study, the nonparametric structural parameters of galaxies in the COSMOS field are estimated using the {\tt\string statmorph\_csst} code, which requires only a few hours of computational time. 

Considering that the measurement of nonparametric structural parameters is influenced by S/Ns, previous studies (e.g., \citealt{lotzNewNonparametricApproach2004, Lotz_2006}, \citealt{treuEarlyResultsGLASSJWST2023}, \citealt{yaoEvolutionNonparametricMorphology2023}) have shown that reliable measurements of these parameters can be obtained when the average S/N per pixel ($\langle$S/N$\rangle$) exceeds 2. Thus, we focused our analysis on galaxies with $\langle$S/N$\rangle$ greater than 2 in this part. The detailed method of obtaining $\langle$S/N$\rangle$ is described in Section 4.3.2 of \cite{rodriguez-gomezOpticalMorphologiesGalaxies2019}.

In Figure \ref{fig10}, we present the distribution of four galaxy subclasses in the Gini-$M_{\rm 20}$ (left) and C-A (right) parameter spaces. Since these parameters are widely used in many studies, we do not introduce the definitions of these parameters, which have been carefully described in Section 4 of \cite{rodriguez-gomezOpticalMorphologiesGalaxies2019}. The contours in this figure represent 50\% of the specified galaxies. Different colored diamonds correspond to the median parameter values for different types of galaxies, with error bars indicating the corresponding standard deviations. It is evident from the left panel that galaxies transition from SPHs to IRRs with a gradual decrease in Gini and a simultaneous increase in $M_{\rm 20}$, which aligns with our common understandings of Gini and $M_{\rm 20}$ for different galaxy types. When focusing on the C-A space, numerous studies have shown that more compact galaxies tend to have larger C coefficients, while more symmetric galaxies tend to have smaller A coefficients. From the right panel, it is evident that the more compact and symmetric SPHs exhibit larger C values and smaller A values, while the more diffuse and irregular IRRs exhibit smaller C values and larger A values, although the difference between A values is small. This is consistent with some previous studies (e.g., \citealt{lotzNewNonparametricApproach2004, Lotz_2006, 2008MNRAS.386..909C, conseliceEvolutionGalaxyStructure2014a}). These findings support the consistency of our classification results with nonparametric parameters.

In addition to these commonly used parameters, some less commonly used but also useful parameters are estimated. In Figure \ref{fig11}, we present the distribution of $G_2$ (left panel) and $\Psi$ (right panel) parameters for different types of galaxies. 

The $G_2$ is designed by \cite{rosaGradientPatternAnalysis2018} to distinguish between elliptical and spiral galaxies based on the symmetry of the flux gradient. By denoting the flux at coordinates $(x_i, x_j)$ in an image as $I(x_i, x_j)$, the local gradient vectors $V(i,  j)$ at $(x_i, x_j)$ can be estimated by calculating the first-order partial of $I(x_i, x_j)$, assuming that the distance between adjacent pixels in the same direction is 1. Based on a given tolerance for norm and phase, symmetric pairs are defined as those that are concentric and have the same modulus and a phase shifted by $\pi$. Then, asymmetric vectors $v_k$ can be obtained after removing all the symmetric pairs from the local gradient vectors. Thus, the $G_2$ coefficient is defined as:
\begin{equation}
G_2 = \frac{V_A}{V}\times \left(2-\frac{|\Sigma_{k=1}^{V_A} v_k|}{\Sigma_{k=1}^{V_A}|v_k|}\right)
\end{equation}
where V is the number of local gradient vectors and $V_A$ is the number of asymmetric vectors. $|\Sigma_{k=1}^{V_A} v_k|$ is the modulus of the sum of asymmetric vectors, while $\Sigma_{k=1}^{V_A}|v_k|$ is the sum of the modulus of asymmetric vectors. 

The left panel of Figure \ref{fig11} displays the distribution of $G_2$ for different galaxy subclasses. Similar to Figure \ref{fig9}, the bars at the top indicate the median values of $G_2$ for each subclass. Galaxies exhibit increasing asymmetry in their flux gradient from SPHs to IRRs, with median $G_2$ values of 
0.39, 1.19, 1.68, and 1.80 for SPHs, ETDs, LTDs, and IRRs, respectively. With a sample from SDSS, \cite{rosaGradientPatternAnalysis2018} found that elliptical and spiral galaxies exhibit a clear bimodal distribution in $G_2$, where the $G_2$ value of spiral (elliptical) galaxies is typically larger (smaller) than 1. Assuming that SPHs are elliptical galaxies and LTDs are spiral galaxies, a similar trend has also been seen in this work. However, ETDs exhibit $G_2$ values that span a wide range, possibly due to their combined bulge and disk structures. Therefore, for galaxies at middle redshift (z$\sim$1), $G_2$ coefficient is also available to distinguish between bulge-- and disk--dominated galaxies, but for galaxies with interim morphologies, it should be treated more carefully.

Furthermore, $\Psi$ was introduced by \cite{lawPhysicalNatureRest2007} to describe how the light distribution of galaxies can be decomposed into apparent components, similar to $M_{\rm 20}$. In general, galaxies that appear more clumpy tend to have larger $\Psi$ values.
By using flux as a proxy for ``mass'', the ``potential energy'' of the observed flux distribution is defined as:
\begin{equation}
\psi_{actual} = \sum_i \sum_j \frac{X_i X_j}{r_{ij}}
\end{equation}
where $X_i$ and  $X_j$ are fluxes of the $i$th and $j$th pixels, and $r_{ij}$ is the distance (in pixels) between the $i$th and $j$th pixels. Then the pixels of raw images are rearranged in a circular configuration that the brightest pixel is at the center, and the flux of the other pixels gradually decreases with distance, which is considered the most compact configuration. The distance between the $i$th and $j$th pixels on this rearranged map is represented as $r^{'}_{ij}$. Then the ``potential energy'' of this rearranged flux distribution is:
\begin{equation}
\psi_{compact} = \sum_i \sum_j \frac{X_i X_j}{r^{'}_{ij}}.
\end{equation}
The $\Psi$ coefficient is then defined as:
\begin{equation}
\Psi = 100 \log_{10}\frac{\psi_{compact}}{\psi_{actual}}.
\end{equation}

The distribution of the $\Psi$ parameter is displayed in the right panel of Figure \ref{fig11}. There is an increasing trend of $\Psi$ from SPHs to IRRs. The median $\Psi$ values are 
0.53, 1.06, 1.34, and 1.72 for SPHs, ETDs, LTDs, and IRRs, respectively. Considering that IRRs are much more clumpy than SPHs, this figure demonstrates that our classification is consistent with $\Psi$.

\subsection{Test of Physical Properties} \label{subsec:physical properties}

Many studies have already shown that there is a close connection between galaxy morphologies and their physical properties. It is now widely agreed that as the stellar mass increases, galaxies become more bulge--dominated (e.g., \citealt{chengOptimizingAutomaticMorphological2020, duEvolutionaryPathwaysDisk2021}), and galaxies at higher redshift are more likely irregular (e.g., \citealt{tohillQuantifyingNonparametricStructure2021, kartaltepeCEERSKeyPaper2023}). In this section, we also select galaxies with stellar masses larger than $10^9M_{\odot}$ to study the relationship between galaxy morphologies and their physical properties.

In Figure \ref{fig12}, the number fractions of galaxies for each morphology class are presented as a function of stellar mass at three different redshift bins (panel a: $0.2<z<0.6$; panel b: $0.6 < z < 0.9$; and panel c: $0.9<z<1.2$). The symbols, including diamonds, squares, stars, and triangles, represent the fractions of SPHs, ETDs, LTDs, and IRRs, respectively, within different stellar mass bins.

In each panel, it is evident that as the stellar mass of galaxies increases, the proportions of SPHs and ETDs gradually rise, while the proportions of LTDs and IRRs decrease. Similar results were also shown in the work of \cite{2013MNRAS.428.1715H}, where at the high-mass end, galaxies are dominated by early-type morphology, while at the low-mass end, galaxies are dominated by late-type morphology. Combining with our earlier discussion (Section \ref{subsec:class_result}) where both SPHs and ETDs exhibit a nuclear bulge structure, while LTDs and IRRs do not have such a structure, this trend suggests that an increase in galaxy stellar mass is accompanied by an increase in bulge dominance.

Comparing the results across different panels, it's also clear that with increasing redshift, the proportion of IRRs continues to increase, while the proportion of SPHs decreases, especially for galaxies with larger $M_{\ast}$. \cite{2013MNRAS.428.1460B} found a similar trend, with the proportions of SPHs was approximately 40\% for massive galaxies at $z\sim 1$, but this proportion increased to 60\% at $z\sim 0.2$. This might indicate that over cosmic time, galaxies tend to evolve toward more regular morphologies. However, at the lower mass end, the change in the ratio of IRRs to SPHs is less pronounced, with SPHs comprising approximately 10\% and IRRs approximately 30\%. This difference could be attributed to lower-mass galaxies being more susceptible to perturbations, such as tidal interactions, ram pressure, and stellar feedback, which can lead to a more irregular appearance. In summary, these results underscore the complex interplay between galaxy morphology, stellar mass, and redshift.

\section{Summary} \label{sec:summary}
In this work, utilizing the galaxy morphology classification algorithm ({\tt\string USmorph}) that combines UML and SML methods, as proposed by \cite{zhouAutomaticMorphologicalClassification2022} and \cite{fangAutomaticClassificationGalaxy2023}, we classify nearly 100,000 galaxies in the COSMOS field into five categories: SPHs (15,200), ETDs (17,369), LTDs (21,143), IRRs (28,965), and UNCs (17,129) using HST/ACS I-band images. In visual inspection,  SPHs exhibit a clear bulge-dominated structure, while LTDs display a clear disk structure. ETDs show a disk structure but have bulge components as well. IRRs include galaxies with irregular structures or merger evidence, while galaxies that could not be confidently identified due to poor S/Ns are classified as UNCs.

Furthermore, we estimate the morphological parameters of these galaxies and find that the relationship between our classification results and galaxy morphological parameters for massive galaxies is consistent with our expectations. In brief, as galaxies transition from IRRs to SPHs, their S\'{e}rsic index gradually increases, and their effective radius decreases. When considering nonparametric morphologies, more compact galaxies (e.g., SPHs) exhibit larger Gini and C coefficients, while more diffuse galaxies (e.g., IRRs) exhibit larger $M_{20}$, A, and $\Psi$ coefficients. In addition, $G_2$ has also been proven to effectively distinguish SPHs from LTDs. Moreover, the relationship between galaxy morphology and their physical properties is investigated using our classification results. A clear relationship was observed between the morphology of galaxies and their stellar mass and redshift. Relevant information is catalogued in the electronic version of the article, and we show a part of the catalog in Table \ref{tab2}.

The forthcoming CSST will be launched in 2024. With a 2-meter aperture and a large field of view, CSST is planned to conduct a 15,000 $\rm deg^2$ multi-band deep field survey with an expected $5\sigma$ depth of $\rm r=26.0$ mag and a 400 $\rm deg^2$ ultra-deep field survey with an expected $5\sigma$ limiting magnitude of $\rm r=27.2$ mag \citep{2011SSPMA..41.1441Z, 2018cosp...42E3821Z}. This will provide us with a large number of high-resolution images. Meanwhile, the {\tt\string USmorph} algorithm can classify about 30,000 galaxies per hour, which could meet the requirements of the CSST surveys. This will help us better utilize future CSST image data for studying galaxy morphology.

\begin{acknowledgements}
This work is supported by the Strategic Priority Research Program of Chinese Academy of Sciences (Grant No. XDB 41000000), the National Science Foundation of China (NSFC, Grant No. 12233008, 11973038), the China Manned Space Project (No. CMS-CSST-2021-A07) and the Cyrus Chun Ying Tang Foundations.
Z.S.L. acknowledges the support from Hong Kong Innovation and Technology Fund through the Research Talent Hub program (GSP028).
Y.Z.G. acknowledges support from the China Postdoctoral Science Foundation funded project (2020M681281). 
C.C.Z. acknowledges the support from National Natural Science Foundation of China (NSFC, Grant No. 62106033).
S.B. acknowledges the support from Scientific Research Fund Project of Yunnan Provincial Department of Education (2023Y1040). 
\end{acknowledgements}

\bibliography{ref}
\bibliographystyle{aasjournal}

\movetabledown=20mm  
\begin{rotatetable*}
\begin{deluxetable*}{cccccccccccccccccccccccccccc}
\centerwidetable
\tablecaption{Part of The Finall Catalogue \label{tab2}}
\tablehead{\colhead{ID} & \colhead{R.A.} & \colhead{DEC.} & \colhead{$I_{\rm mag}$} & \colhead{z} & \colhead{$M_{\ast}$} & \colhead{$r_e$} & \colhead{n} & \colhead{$\rm flag1$} & \colhead{C} & \colhead{A} & \colhead{S} & \colhead{G} & \colhead{$M_{20}$} & \colhead{M} & \colhead{I} & \colhead{D} & \colhead{$\Psi$} & \colhead{$G_{2}$} & \colhead{$\rm flag2$} & \colhead{Type} \\
 & (deg) & (deg) & mag &  & ($\log M_{\odot}$) & (kpc) & & & & & & &  & & &  & & & & & & \\ 
(1) & (2) & (3) & (4) & (5) & (6) & (7) & (8) & (9) & (10) & (11) & (12) & (13) & (14) & (15) & (16) & (17) & (18) & (19) & (20) & (21) \\}
\startdata  
1 & 150.71249 & 1.79211 & 24.83 & 0.78 & 8.52 & 0.95 & 0.83 & 2 & 2.46 & -0.17 & 0.03 & 0.46 & -1.59 & 0.02 & 0.00 & 0.13 & 1.90 & 1.96 & 0 & IRR \\
2 & 150.71818 & 1.79273 & 24.60 & 0.97 & 10.44 & 0.39 & 8.00 & 2 & 2.81 & -0.01 & 0.00 & 0.48 & -1.78 & 0.05 & 0.00 & 0.06 & 1.06 & 1.88 & 0 & LTD \\
3 & 150.72869 & 2.54382 & 24.66 & 0.36 & 7.94 & 1.42 & 1.92 & 2 & 3.15 & -0.14 & -0.07 & 0.51 & -1.63 & 0.27 & 0.40 & 0.28 & 2.38 & 1.79 & 0 & LTD \\
4 & 149.63794 & 2.37729 & 24.06 & 0.89 & 9.09 & 3.14 & 0.20 & 2 & 1.89 & -0.12 & 0.02 & 0.41 & -0.83 & 0.24 & 0.76 & 0.29 & 8.60 & 1.78 & 0 & IRR \\
5 & 150.71562 & 1.79362 & 21.31 & 0.39 & 9.66 & 4.36 & 0.72 & 2 & 2.70 & 0.01 & 0.03 & 0.46 & -1.49 & 0.15 & 0.46 & 0.06 & 0.00 & 1.89 & 0 & IRR \\
6 & 150.71684 & 1.79427 & 23.60 & 0.65 & 8.99 & 1.03 & 1.95 & 2 & 2.90 & -0.12 & 0.02 & 0.48 & -1.79 & 0.02 & 0.00 & 0.04 & 0.52 & 1.78 & 0 & ETD \\
7 & 150.47448 & 2.33798 & 24.42 & 0.95 & 8.90 & 2.41 & 1.38 & 2 & 2.53 & -0.35 & 0.02 & 0.38 & -1.22 & 0.55 & 0.14 & 0.15 & 2.21 & 1.95 & 0 & IRR \\
8 & 150.71614 & 1.79478 & 24.79 & 0.66 & 8.22 & 0.40 & 5.12 & 2 & 3.18 & 0.01 & 0.05 & 0.55 & -1.75 & 0.04 & 0.00 & 0.09 & 1.12 & 1.78 & 0 & LTD \\
9 & 150.49108 & 1.87574 & 22.54 & 0.45 & 9.80 & 1.81 & 8.00 & 2 & 3.53 & 0.05 & 0.00 & 0.59 & -2.00 & 0.03 & 0.00 & 0.03 & 0.30 & 0.47 & 0 & SPH \\
10 & 150.35723 & 2.29304 & 23.95 & 0.77 & 8.04 & 1.78 & 6.20 & 2 & 3.55 & 0.02 & -0.02 & 0.60 & -1.86 & 0.04 & 0.00 & 0.12 & 1.67 & 1.66 & 0 & SPH \\
11 & 149.97418 & 1.70920 & 23.81 & 1.00 & 8.90 & 2.20 & 1.10 & 2 & 2.71 & 0.16 & 0.04 & 0.50 & -1.34 & 0.23 & 0.27 & 0.19 & 3.65 & 1.81 & 0 & LTD \\
12 & 150.71988 & 2.54496 & 24.19 & 0.66 & 8.70 & 0.82 & 0.87 & 2 & 2.46 & 0.08 & 0.05 & 0.50 & -1.53 & 0.01 & 0.00 & 0.16 & 0.61 & 1.88 & 0 & SPH \\
13 & 150.71874 & 2.54548 & 21.06 & 0.61 & 10.32 & 5.47 & 1.90 & 2 & 3.14 & 0.11 & 0.03 & 0.56 & -1.94 & 0.02 & 0.10 & 0.03 & 0.00 & 0.93 & 0 & ETD \\
14 & 150.49124 & 1.87773 & 23.29 & 0.96 & 9.66 & 2.76 & 1.29 & 2 & 2.73 & 0.09 & 0.01 & 0.45 & -1.64 & 0.06 & 0.14 & 0.11 & 1.67 & 1.86 & 0 & IRR \\
15 & 149.63540 & 2.37918 & 22.78 & 0.92 & 10.45 & 1.45 & 2.66 & 2 & 3.18 & -0.01 & 0.01 & 0.55 & -1.89 & 0.01 & 0.00 & 0.01 & 1.40 & 1.14 & 0 & SPH \\
16 & 149.96909 & 1.70877 & 24.71 & 0.93 & 9.09 & 3.54 & 1.29 & 2 & 2.96 & 0.01 & 0.03 & 0.45 & -1.55 & 0.16 & 0.36 & 0.20 & 6.72 & 1.84 & 0 & IRR \\
17 & 150.72626 & 2.54642 & 22.00 & 0.36 & 9.39 & 1.67 & 1.50 & 2 & 3.02 & 0.07 & 0.02 & 0.53 & -1.84 & 0.01 & 0.01 & 0.04 & 1.12 & 1.32 & 0 & ETD \\
18 & 149.63316 & 2.37972 & 24.66 & 1.09 & 10.22 & 7.32 & 6.57 & 2 & 3.35 & -0.09 & 0.03 & 0.48 & -1.70 & 0.12 & 0.08 & 0.14 & 3.60 & 1.73 & 0 & ETD \\
19 & 149.79397 & 2.37765 & 21.18 & 0.43 & 10.19 & 4.24 & 1.06 & 2 & 2.72 & 0.04 & 0.02 & 0.48 & -1.77 & 0.02 & 0.10 & 0.04 & nan & 1.72 & 0 & LTD \\
20 & 150.72653 & 1.79572 & 22.05 & 0.82 & 9.34 & 2.95 & 4.35 & 2 & 3.91 & 0.24 & 0.02 & 0.57 & -1.88 & 0.42 & 0.17 & 0.02 & 3.53 & 0.86 & 0 & SPH \\
\enddata
\tablecomments{(1) Sequential number identifier; (2) R.A. expressed in decimal degrees; (3) decl. expressed in decimal degrees; (4), (5), and (6) are magnitude in I band, redshift, and stellar mass from \cite{weaverCOSMOS2020PanchromaticView2022}, respectively; (7) effective radius; (8) S\'{e}rsic index; (9) flag of single S\'{e}rsic fitting, 2 represents a good fitting; (10), (11), (12), (13), (14), (15), (16), (17), (18), and (19) represent C, A, S, Gini, $M_{\rm 20}$, M, I, D, $\Psi$, and $G_2$ coefficient, respectively, (20) flag of the nonparametric morphological measurements, 0 represents a good measurement; (21) galaxy morphological type.}
\end{deluxetable*}
\end{rotatetable*}

\end{document}